# CRITICAL COMPARISONS ON DEEP LEARNING APPROACHES FOR FOREIGN EXCHANGE RATE PREDICTION

Zhu Bangyuan

School of Electrical & Electronic Engineering
Nayang Technological University
Nanyang Avenue, Singapore 639798
e-mail: zhub0005@ntu.edu.sg

**Keywords:** Deep learning · Foreign exchange rate · Recurrent neural network · LSTM network · BP neural network

## Abstract

The correct analysis and forecast of the exchange rate have always been of great significance to the formulation of relevant economic and financial policies and the avoidance of foreign exchange risks by enterprises. The foreign exchange market is a dynamic market with nonlinear changes, and exchange rate fluctuations have very significant nonlinear and historical dependence characteristics. For foreign exchange, its price at a certain point in the future is closely related to historical data and has a strong time correlation. To study prediction tasks for such time-dependent data, consider choosing recurrent neural networks as a research direction.

In a real market environment, the price prediction model needs to be updated in real-time according to the data obtained by the system to ensure the accuracy of the prediction. In order to improve the user experience of the system, the price prediction function needs to use the fastest training model and the model prediction fitting effect of the best network as a predictive model. We conduct research on the basic theories of RNN, LSTM, and BP neural networks, analyse their respective characteristics, and discuss their advantages and disadvantages, so as to provide a reference for the selection of price prediction models.

## 1 Introduction

The foreign exchange market is a multi-variable nonlinear system, and there are many and complicated factors affecting the foreign exchange rate [1]. The tools people use in the process of analysing and predicting foreign exchange rates have undergone a series of changes. At the beginning, people tried to find the laws of foreign exchange rates through linear models and time series models, but they did not find many valuable conclusions during the research process. Later, with the further development of research, scholars noticed the advantages of the neural network model in dealing with nonlinear characteristic systems, so some scholars began to apply neural networks to the field of foreign exchange and other similar fields with nonlinear characteristics [2]. Although the neural network has a unique effect in dealing with nonlinear problems, the foreign exchange time series has quite significant historical dependence characteristics, and the traditional neural network model cannot capture and analyse this characteristic.

Scholars Chortareas made fluctuation predictions on the euro exchange rate data and proposed that if high-frequency data and long-term memory dimensions can be added to the prediction, the prediction effect can be greatly improved [3]. On the other hand, some scholars have long demonstrated that the time series forecasting method has a shortcoming: the forecasting effect of the model is better when the time interval is shorter. However, as the time interval increases and the data samples decrease, the prediction error of the model will also increase.

At the beginning of the 20th century, some scholars used neural networks to predict exchange rates. Scholar Refense studies using artificial networks to predict the foreign exchange market [4]. After machine learning training models and verification on the test set, artificial neural networks have obvious advantages in fitting foreign exchange rates compared to traditional time series methods. Scholars Yao used the BP neural network to predict the foreign exchange rate [5], and compared the prediction effect of the BP neural network and the traditional ARIMA forecasting model on the basis of a series of currency pair exchange rates. It is found that compared with the traditional ARIMA prediction model, the prediction effect of BP neural network has a huge improvement.

## 2 Neural network architectures

### 2.1 Recurrent neural network

The main purpose of the recurrent neural network RNN is to process and predict sequence data, emphasizing the order of time, and believes that the subsequent value is determined by the probability of the former and some other parameters. This view is consistent with the change of foreign exchange prices.

Compared with the simple feedforward neural network structure, the biggest difference of the cyclic neural network is that the output of the cyclic neural network can not only propagate to the next layer, but also can be passed to the next moment of the same layer. For a certain neuron node in the recurrent network, its internal operation data not only includes the output of the previous layer, but also includes the output of the same layer at the previous moment. The structure is shown below:

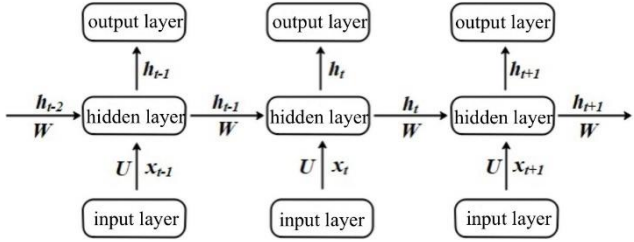

Figure1: RNN neural network structure.

In the figure, $U$ and $W$ are the weights of $x$ and $y$ respectively, so the operations in hidden layer nodes can be expressed as follows:

$$h_t = f(U \times x_t + W \times h_{t-1} + b) \quad (1)$$

where $f$ is the activation function and $b$ is the bias. It can be seen that when calculating the output $h_t$ at the $t$th moment, the output state $h_{t-1}$ at the previous moment also participates in the calculation, and for $h_{t-1}$:

$$h_{t-1} = f(U \times x_{t-1} + W \times h_{t-2} + b) \quad (2)$$

Combine (1) with (2), we can get:

$$h_t = f(U \times x_t + f(U \times x_{t-1} + W \times h_{t-2} + b) + b) \quad (3)$$

In this way, the output at the $t$th moment is associated with the output at the previous moment, and there is a previous "memory".

Due to the problems of gradient explosion and gradient disappearance in RNN itself, when RNN is applied to foreign exchange price prediction, the complexity of the network will affect its prediction effect. In order to meet the accuracy and efficiency of price prediction, we need to verify it through experiments.

### 2.2 Long short-term memory network

Long short-term memory network, also known as LSTM, is one of the special RNN structure. It was first introduced by Hochreiter & Schmidhuber in 1997 [4], and recently improved and promoted by Grave [5].

In order to solve the problem of long-Term dependencies, researchers have proposed many solutions, such as ESN (Echo State Network), adding leaky units (Leaky Units) and so on. The most successful and widely used one is the threshold RNN (Gated RNN), and LSTM is the most famous one in the threshold RNN. The leaky unit allows RNN to accumulate long-term connections between distant nodes by designing the weight coefficients between connections. The threshold RNN generalizes this idea, allowing the coefficient to be changed at different times, and allowing the network to forget the current accumulated information.

The main character of LSTM is that by increasing the input threshold, forgetting threshold and output threshold, the weight of the self-loop is changed. In this way, when the model parameters are fixed, the integration scale at different moments can be changed dynamically, thus avoiding the problem of gradient disappearance or gradient expansion.

According to the structure of the LSTM network, the calculation formula of each LSTM unit is shown below, where $F_t$ represents the forgetting threshold, $i_t$ represents the input threshold, $\tilde{C}_t$ represents the cell state at the previous moment, and $C_t$ represents the cell state (here is where the cycle occurs), $o_t$ represents the output threshold, $H_t$ represents the output of the current unit, and $H_{t-1}$ represents the output of the unit at the previous moment.

$$f(t) = \sigma(W_f \cdot [h_{t-1}, x_t] + b_f) \quad (4)$$

$$i_t = \sigma(W_i \cdot [h_{t-1}, x_t] + b_i) \quad (5)$$

$$\tilde{C}_t = tanh(W_C \cdot [h_{t-1}, x_t] + b_C) \quad (6)$$

$$C_t = f_t * C_{t-1} + i_t * \tilde{C}_t \quad (7)$$

$$o_t = \sigma(W_o \cdot [h_{t-1}, x_t] + b_o) \quad (8)$$

$$h_t = o_t * tanh(C_t) \quad (9)$$

### 2.3 Back propagation neural network

The classic BP neural network usually consists of three layers: input layer, hidden layer and output layer. Usually, the number of neurons in the input layer is related to the number of features, the number of output layers is the same as the number of categories, and the number of layers and neurons in the hidden layer can be customized.

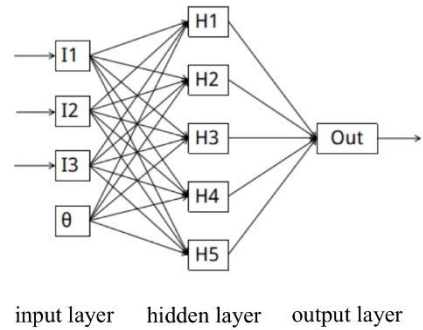

Figure2: BP neural network structure.

The functional relationship between the output and input of each neuron in the hidden layer and output layer is:

$$I_j = \sum_i W_{ij} O_j \quad (10)$$

$$O_j = sigmod(I_l) = \frac{1}{1+e^{-I_l}} \quad (11)$$

Where $W_{ij}$ represents the weight of the connection between neuron $i$ and neuron $j$, $O_j$ represents the output of neuron $j$, and sigmod is a special function used to map any real number to the (0, 1) interval. The sigmod function above is called the activation function of neurons. In addition to the sigmod function $\frac{1}{1+e^{-I_l}}$, tanh and ReLU functions are commonly used [6].

As a supervised learning algorithm, the training process of BP neural network is the process of comparing the predicted value obtained by feedforward with the reference value, and adjusting the connection weight $W_{ij}$ according to the error.

The training process is called the backpropagation process, and the data flow is just the opposite of the feedforward process.

First, the connection weight $W_{ij}$ is randomly initialized, and a feedforward process is performed on a certain training sample to obtain the output of each neuron. Calculate the error of the output layer:

$$E_j = sigmod'(O_j) * (T_j - O_j) = O_j(1 - O_j)(T_j - O_j) \quad (12)$$

Where $E_j$ represents the error of neuron $j$, $O_j$ represents the output of neuron $j$, $T_j$ represents the reference output of the current training sample, and $sigmod'$ is the first derivative of the above sigmod function.

Calculate the hidden layer error:

$$E_j = sigmod'(O_j) * \sum_k E_k W_{jk} = O_j(1 - O_j) \sum_k E_k W_{jk} \quad (13)$$

There is no reference value in the output of the hidden layer, and the weighted sum of the error of the next layer is used instead of $(T_j - O_j)$.

After calculating the error, $W_{ij}$ and $\theta_j$ can be updated:

$$W_{ij} = W_{ij} + \lambda E_j O_i \quad (14)$$

Where $\lambda$ is a parameter called the learning rate, which generally takes a value on the interval $(0, 0.1)$.

## 3 Data collection

We choose the minute-level closing price data of the Australian dollar, Swiss franc, British pound, Canadian dollar, and Euro against the US dollar published by myfxbook as our basic data. The exchange rate data released by the platform is relatively complete and accurate to the minute level, which meets the data analysis needs of investors and researchers in the international market for short-term high-frequency exchange rate series research. This article uses the three months from January 4, 2017 to March 31, 2017 as our observation period.

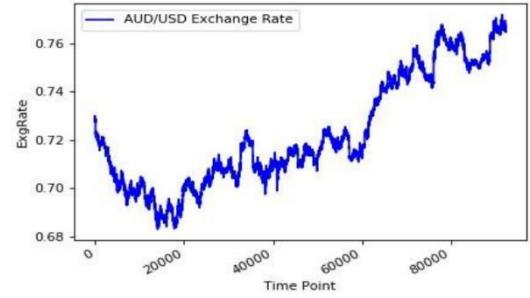

Figure3: AUD/USD exchange rate

| mean | std | min | Q1 | Q2 | Q3 | max |
|---|---|---|---|---|---|---|
| 0.722 | 0.023 | 0.023 | 0.705 | 0.716 | 0.746 | 0.772 |

Table1: AUD/USD closing price statistics

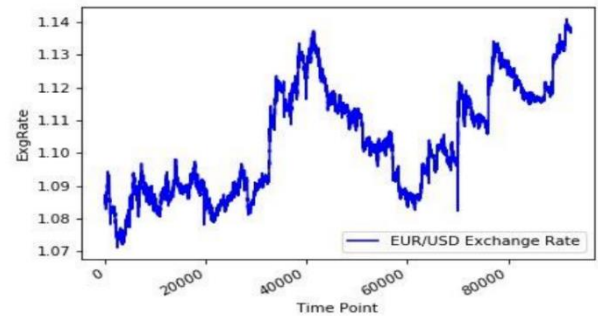

Figure4: EUR/USD exchange rate

| mean | std | min | Q1 | Q2 | Q3 | max |
|---|---|---|---|---|---|---|
| 1.103 | 0.016 | 1.071 | 1.088 | 1.101 | 1.117 | 1.141 |

Table2: EUR/USD closing price statistics

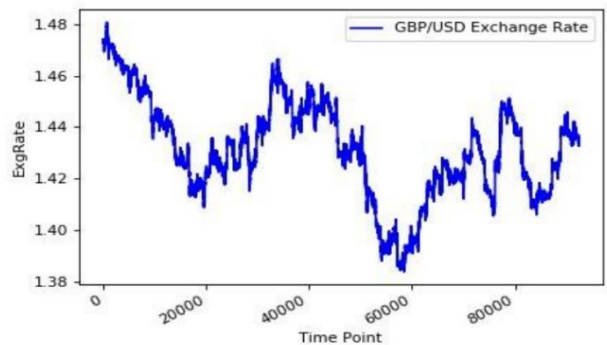

Figure4: GBP/USD exchange rate

| mean | std | min | Q1 | Q2 | Q3 | max |
|---|---|---|---|---|---|---|

| 1.239 | 0.015 | 1.199 | 1.227 | 1.243 | 1.25 | 1.27 |

Table3: GBP/USD closing price statistics

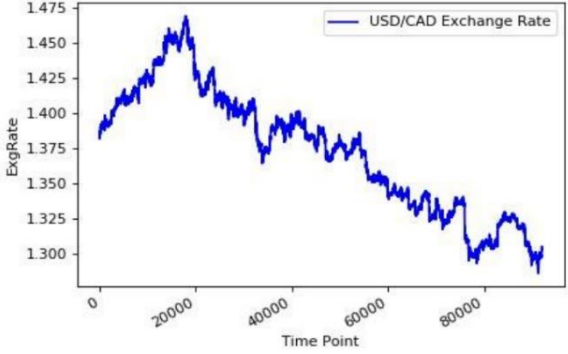

Figure5: USD/CAD exchange rate

| mean | std | min | Q1 | Q2 | Q3 | max |
|---|---|---|---|---|---|---|
| 1.373 | 0.044 | 1.286 | 1.334 | 1.378 | 1.407 | 1.469 |

Table4: USD/CAD closing price statistics

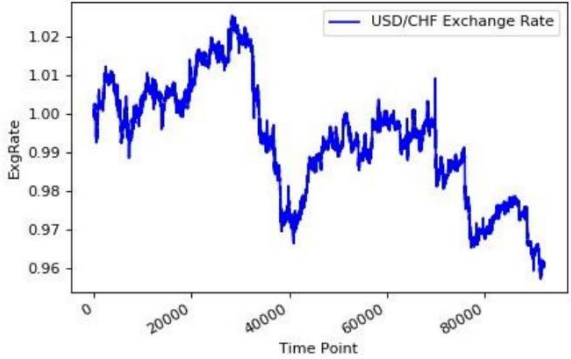

Figure6: USD/CAD exchange rate

| mean | std | min | Q1 | Q2 | Q3 | max |
|---|---|---|---|---|---|---|
| 0.993 | 0.015 | 0.957 | 0.984 | 0.994 | 1.004 | 1.025 |

Table5: USD/CHF closing price statistics

## 3 Data modelling

In order to facilitate the modelling of predictive models, it is often necessary to convert data into time series to construct samples. Taking the time lag as $\Delta t$ time units as an example, the data of $t + \Delta t$ is used as the feature as the input value of the model, and the data of $t'$ at the next moment is the label as the predicted value, thus forming a time series window. Then every time the window rolls backward by one time unit, such as from $t$ to $t + 1$, a new time series record will be formed. In this record, the feature of the model is the data in $t + \Delta t + 1$, and the label is the data in $t' + 1$. According to this mechanism, the time series record is constructed until the window rolls to the last time point.

## 4 Model parameters

In order to accelerate the optimization of model parameters, accelerate convergence and extract the influence of dimensions on the prediction effect, it is necessary to normalize the sequence data. The minmax function is used to transform the sequence data and inversely transform the prediction results, so as to transform the numerical mapping of all sequences into the (0,1) interval. The formula for the normalized transformation process is:

$$x_i' = \frac{x_i - x_{min}}{x_{max} - x_{min}} \quad (15)$$

Among them, $x_i$, $x_i'$ represent the exchange rate value and normalized value at the $i$-th time point, respectively, $x_{min}$, $x_{max}$ are the minimum and maximum value of the exchange rate at all time points. The formula for the denormalization process is:

$$\hat{y}_l = y_i^* * (x_{max} - x_{min}) + x_{min} \quad (16)$$

Among them, $y_i^*$ is the predicted value after model training, and $\hat{y}_l$ is the denormalized value of the predicted value.

As for relevant network parameters of the LSTM neural network model, this paper draws on the neural network structure of Simplex LSTM used in Li and Cao (2018) [7] and improves it. The LSTM neural network used is divided into 8 layers, including 6 layers of LSTM hidden layers, 1 layer of LSTM input layer and 1 layer of fully connected output layer. The number of input neurons is 10, and the number of output neurons is 1, that is, the exchange rate closing price vector at 10 time points is used as a feature, and the exchange rate value at the next time point is a label to train the model. The number of neurons in each layer of the other four hidden layers is 128. The LSTM input layer activation function uses the modified linear unit function relu, the hidden layer neuron activation function uses the tanh function, and the output layer activation function uses the linear function. In addition, this paper refers to N Srivastava et al. (2014) to introduce dropout in the LSTM neural network to reduce the overfitting of the model [8], and set the dropout ratio to 0.1. The calculation forms of these activation functions are shown in the table:

relu $\quad f(x) = max(0, x) \quad (17)$

tanh $\quad f(x) = \frac{e^x - e^{(-x)}}{e^x + e^{(-x)}} \quad (18)$

linear $\quad f(x) = x \quad (19)$

The BP neural network used in this paper has the same number of layers as 8 layers, the number of neurons in the hidden layer is also 128, and the activation function uses tanh.

## 5 Model performance evaluation

In the process of the exchange rate forecasting experiment, in order to objectively reflect the forecasting performance of the model, this paper uses three different indicators to evaluate the forecasting effect of the model, including mean square error (RMSE), mean absolute error (MAE) and mean absolute percentage error (MAPE). The formula is following:

$$RMSE = \sqrt{\frac{1}{n}\sum_{i=1}^{n}(y_i - y_i^*)^2} \qquad (20)$$

$$MAE = \frac{1}{n}\sum_{i=1}^{n}|y_i - y_i^*| \qquad (21)$$

$$MAPE = \frac{1}{n}\sum_{i=1}^{n}\frac{|y_i - y_i^*|}{y_i^*} \qquad (22)$$

Among them, $n$ is the number of samples, $y_i$, $y_i^*$ are the predicted value and real value of the $i$th sample, respectively. MAE evaluates the absolute difference between the true value and the predicted value. This metric is more robust to large errors than other metrics. RMSE reflects the square error of two values, is more sensitive to large errors, and is more likely to amplify the impact of errors. MAPE is the ratio between the error and the true value and is a relative error function. For a small error, it may have a large impact in the interval of the overall small variable value, while a large error has a small impact in the interval of high variable values. These three metrics reflect the performance of the model from different perspectives.

## 6 Prediction Result

Through the data we obtained, data processing is performed on these six exchange rate data, and the exchange rate samples at the 5-minute time node are respectively extracted as the experimental data for our prediction. Then, for the exchange rate series data of all currency pairs, the training set and the test set are divided according to the ratio of 8:2.

In order to objectively evaluate the prediction effect of these models, we calculate the three prediction error indicators on the verification set as shown in the table:

| error index | MAE(1e-3) | | | RMSE(1e-3) | | | MAPE(%) | | |
|---|---|---|---|---|---|---|---|---|---|
| model | LSTM | BP | RNN | LSTM | BP | RNN | LSTM | BP | RNN |
| AUD/USD | 0.55 | 0.65 | 7.1 | 0.81 | 0.8 | 9.1 | 0.07 | 0.08 | 0.9 |
| EUR/USD | 1.5 | 1.6 | 11.6 | 1.86 | 1.89 | 12.6 | 0.13 | 0.14 | 1.02 |
| GBP/USD | 2.2 | 4.41 | 4.72 | 2.44 | 5.21 | 5.63 | 0.15 | 0.31 | 0.32 |
| USD/CAD | 6 | 15.5 | 45.6 | 6.97 | 16 | 46.4 | 0.47 | 1.22 | 3.6 |
| USD/CHF | 1 | 5.2 | 7.1 | 1.22 | 5.53 | 7.8 | 0.1 | 0.54 | 0.74 |

Table6: Multi-currency exchange rate forecast error table

## 7 Conclusion

In terms of high-frequency exchange rate time series forecasting, this paper designs six foreign exchange rate forecasting experiments at the minute granularity level, and compares the forecasting performance of the LSTM model, RNN, and BP neural network at different time granularities. A number of indicators show that the LSTM model is significantly better than the other two models in terms of short-term foreign exchange rate forecasting.